\documentclass[11pt,twoside]{article}

\usepackage{asp2006}
\usepackage{epsf}
\usepackage{lscape}

\begin{document}
\title{The Coolest Stars in the Clouds: Unusual Red Supergiants in the Magellanic Clouds}   
\author{Emily M. Levesque}   
\affil{Institute for Astronomy, University of Hawaii}    
\author{Philip Massey}
\affil{Lowell Observatory}
\author{K. A. G. Olsen}
\affil{National Optical Astronomy Observatories}
\author{Bertrand Plez}
\affil{GRAAL, Universit\'{e} de Montpellier II, CNRS}

\begin{abstract} 
Red supergiants (RSGs) are a He-burning phase in the evolution of moderately high mass stars (10-25 solar masses).
The evolution of these stars, particularly at low metallicities, is still poorly understood. The latest-type RSGs in
the Magellanic Clouds are cooler than the current evolutionary tracks allow, occupying the region to the right of the
Hayashi limit where stars are no longer in hydrostatic equilibrium. We have discovered four Magellanic Cloud RSGs in this region
that display remarkably similar unusual behavior. All of the, show considerable variations in their $V$ magnitudes and
effective temperatures (and spectral types). Two of these stars, HV 11423 and [M2002] SMC 055188, have been observed
in an M4.5 I state, considerably later and cooler than any other supergiant in the SMC. These stars suffer dramatic
physical changes on timescales of months - when they are at their warmest they are also brighter, more luminous,
and show an increased amount of extinction. This variable extinction is characteristic of the effects of circumstellar
dust, and can be connected with sporadic dust production from these stars in their coolest states. We suggest that
these unusual properties are indicative of an unstable (and short-lived) evolutionary phase not previously associated
with RSGs, and consider the implications such behavior could have for our understanding of the latest stages of massive
star evolution in low-metallicity environments.
\end{abstract}


\section{Introduction}   
Red supergiants (RSGs) are a He-burning phase in the evolution of moderately high mass
stars (10-25$M_{\odot}$). Until recently, the location of RSGs on the H-R diagram was not accurately reproduced by stellar
evolutionary theory, with the stars characterized as too cool and too luminous to agree
with the predictions of the evolutionary tracks. This places them at odds with the restrictions imposed by the Hayashi
limit, which denotes the largest radius a star of a given mass can have while still remaining
in hydrostatic equilibrium \citep{ha1961}. Stars which are cooler
than the tracks allow inhabit the region to the right of this limit on the H-R diagram, suggesting
that they are not in hydrostatic equilibrium.

There are many explanations as to why RSGs might occupy this ``forbidden'' region of the
H-R diagram. RSGs present significant challenges to evolutionary theory. The velocities
of the convective layers are nearly sonic and even supersonic in the atmospheric layers,
giving rise to shocks \citep*{fr2002} and invalidating mixing-length assumptions.
There is poor knowledge of RSG molecular opacities. Finally, the highly-extended atmospheres
of these stars differ from the plane-parallel geometry assumption of the evolutionary models.
As a result of these uncertainties, the error could lie in the position of the evolutionary tracks on
the H-R diagram.
Another possibility is that the ``observed'' location of RSGs in the H-R diagram is incorrect
as a result of inaccurate effective temperatures -
past effective temperature scales for red supergiants have been based on temperatures from
broadband colors, which are highly sensitive to surface gravity and the adopted reddening
for the sample \citep{phil1998,me2005}. Finally, it is possible that some stars
might truly occupy this region, suggesting that they are unusual stars in a unique
and unstable evolutionary state.

\section{Milky Way Red Supergiants}
To shed some light on this debate, \citet{me2005} redetermined the effective temperature scale for
RSGs in the Milky Way. We fit moderate-resolution optical spectrophotometry of 74
Galactic RSGs with the new generation of MARCS stellar atmosphere models \citep{gu2003,pl1992},
which include a much improved treatment of molecular opacities \citep{pl2003,gu2003}. With these models
we were able to use the rich and temperature-sensitive TiO molecular
bands to measure their effective temperatures. The newly derived physical parameters brought
the location of the RSGs into much better agreement with the predictions of stellar
evolutionary theory - there is now excellent agreement
between Milky Way supergiants and the evolutionary tracks. This suggests that, on the whole,
the theoretical models do a good job of accurately reproducing this stage of massive
star evolution in the Milky Way. However, there are occasional exception to this that must
be considered.

\subsection{VY Canis Majoris}
VY Canis Majoris is a peculiar late-type M supergiant in the Milky Way. It has several
physical characteristics unique among its counterparts, mostly related to its unusual
circumstellar environment. It has a large IR excess, making it one of the brightest
5-20 $\mu$m objects in the sky, indicative of a dust shell or disk being heated
by the star \citep{he1970}. Such a dust shell is overwhelmingly present - the RSG is surrounded
by an asymmetric dust-reflection nebula that extends 8''-10'' from the star \citep{mo1999,sm2001,sm2004,hu2005}.
Its mass-loss rate is quite high
for a RSG, about 2 $\times$ 10$^{-4} M_{\odot}$ yr$^{-1}$ \citep{da1994}. The
photometric history of this star dates back to 1801; the star has faded by about 2 mag during
that time but has showed little change in its color \citep{ro1970,ro1971,phil2006}.

This star is often assumed to have extraordinary physical properties \citep{le1996,sm2001,mo2004,hu2005}.
Its assumed effective
temperature for many years was a remarkably cold 2800 K, a value adopted by \citet{le1996}
said to be based on the star's assigned M4 I spectral type (the effective
temperature scale of \citet{dy1974} is cited, but the 2800 K effective tempearature they
report is much cooler than the 3400 K that the Dyck scale associates with M4 supergiants). VY CMa's
luminosity is often cited as approximately $M_{bol}$ = -8.5 to -9.5. With such a high
luminosity and cold temperature, this places VY CMa in a considerably cooler and more
luminous position that current evolutionary tracks allow, inhabiting the previously mentioned
Hayashi forbidden region (Figure 2).. Such a star would have a radius of 1800 to 3000 $R_{\odot}$, making
it far bigger than the largest RSGs described in \citet{me2005} and far outside the limitations of
evolutionary theory.

A proper understanding of this star's physical properties hinges on an accurate determination
of its effective temperature. In \citet{phil2006} we used the same method of model fitting
to redetermine VY CMa's temperature spectroscopically.
A satisfactory fit of $T_{\rm eff}$ = 3650 K was quickly obtained. We reclassified the
star's spectral type as an M2.5 I based on the strength of the TiO bands typically used for
such a classification \citep[$\lambda\lambda$ 6158, 6658, 7054; see][]{ja1990}
- this spectral
type and effective temperature agree excellently with the effective temperature scale for Milky Way
RSGs. Since many earlier studies describe VY CMa as an M3 I to M5 I supergiant \citep{jo1942,wa1958,hu1974}
based on blue (photographic) spectrograms, we also used
the TiO bands further to the blue ($\lambda\lambda$ 4761, 4954, 5167) to assign a spectral
type and temperature, and arrived at a spectral type of M4 I (in good agreement with earlier
studies) and a $T_{\rm eff}$ = 3450 K, a cooler (and poorer overall) temperature that is still
much higher than the commonly-adopted 2800 K temperature of \citet{le1996}.
From these temperatures we calculated a $M_{\rm bol}$ = -7.2 for the red TiO bands, and $M_{\rm bol}$
= -6.9 for the blue TiO bands, as described in \citet{phil2006}.

With these new physical parameters, we found that VY CMa now occupied a much less surprising
region of the H-R diagram (see Figure 1).
The new $T_{\rm eff}$ and $M_{\rm bol}$ derived from spectral fitting place the
star comfortably in agreement with the rightmost reaches of the 15 M$_{\odot}$ evolutionary
tracks.

\begin{figure}[!ht]
\plotone{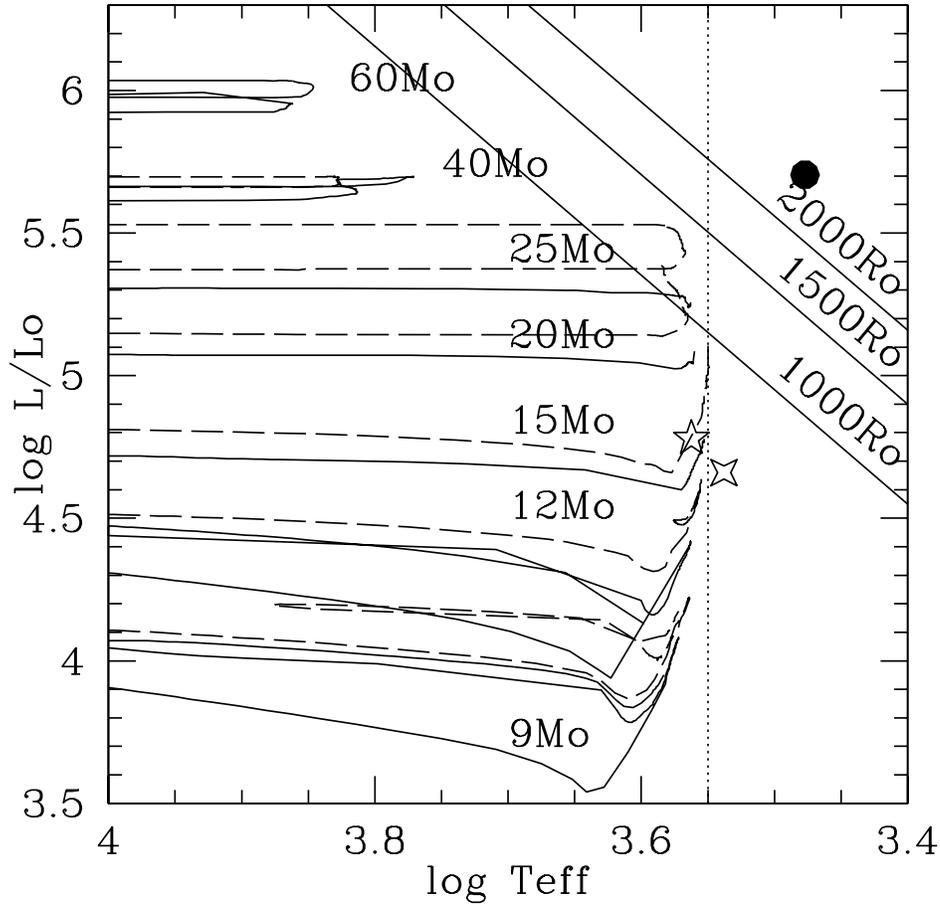}
\caption{Placement of VY CMa in the H-R diagram. The filled circle corresponds
to the parameters of VY CMa from \citet{sm2001}. The stars represent the
results of our work, based on fitting the MARCS models to our
spectrophotometry - the five-pointed star corresponds to the parameters
derived from fitting the red TiO bands, while the four-pointed star
corresponds to the parameters derived from fitting the blue TiO bands. The
evolutionary tracks are from \citet{mey2003} and correspond to models
with no initial rotation (solid lines) and 300 km s$^{-1}$ of initial rotation
(dashed lines). The solid diagonal lines at the upper right represent
lines of constant radii. The Hayashi limit is represented by the dotted
vertical line.}
\end{figure}

However, our method seems to have underestimated the luminosity of the  
star, and Roberta Humphreys (private communication) was kind enough to call
this to our attention.  We can demonstrate the problem like so:

From JHK, \citet{phil2005} were able to determine the properties  
of the dust shell, and found an effective temperature of 760 K and an  
area that is 2155 times that of the star's area.  This leads to an  
interesting contradiction: the luminosity of the dust shell must be  
4x that of the star, or about  $2.3x10^5 L_\odot$.   If the star's  
luminosity is the only source of the heating of the dust, and the  
dust is in thermal/radiative equilibrium with its surrounds, then  
this is very hard to understand.

It is clear that something has caused us to underestimate the luminosity of the star.
One explanation would be if the star's dust had a larger grain size
distribution than normal, and that its extinction was much ``greyer"  
than normal as a result. This would require a lot of extra extinction - about  
1.5-2.0 mag - but would present one way out of this conundrum.

Despite our error in determining the luminosity of the central star in this unusual object,
it should be pointed out that a change in bolometric luminosity on the H-R digram shifts
the location of VY CMa upward on the H-R diagram but does not disrupt its position with respect
to the Hayashi limit, as its temperature still places it solidly at the edge of
higher-mass evolutionary tracks, where it remains in hydrostatic equilibrium. Even if the
proposed upper limit on the bolometric luminosity of $M_{\rm bol}$ = -9.5 is considered, we
can still see that the star would simply move from being a stable highly-evolved
15 M$_{\odot}$ RSG to being a stable highly-evolved 25 M$_{\odot}$ RSG - its maximum
radius would still only about about 1200 R$_{\odot}$ and it would still remain comfortably
in hydrostatic equilibrium at the limits of the Hayashi track.

\section{Magellanic Cloud Red Supergiants}
With this newfound agreement between the Milky Way RSGs and their evolutionary tracks,
we turned out attention to a similar problem
for RSGs in the Magellanic Clouds \citep{me2006}. From \citet{phil2003} we find that again the
evolutionary tracks for the Large and Small Magellanic Clouds (LMC and SMC, respectively)
do not extend to cool enough
temperatures to accomodate the ``observed'' locations of RSGs in the H-R diagram. We applied
the same fitting method described above and in \citet{me2005}, obtaining moderate-resolution
spectrophotometry of 36 LMC RSGs and 37 SMC RSGs using the R-C spectrograph on the CTIO Blanco
4 m telescope in November and December of 2004. We fit these data with the MARCS models to
redetermine the
effective temperature scales for RSGs in the Clouds. The new effective temperature
scales (see Figure 2) brought the MC RSGs
into much better agreement with stellar evolutionary theory. The current agreement in the LMC
is good, while the agreement in the SMC is improved but still not satisfactory, with SMC RSGs
showing a considerably larger spread in effective temperatures across a given luminosity
than their LMC counterparts. Such a spread is expected, however, due to the larger effects of
rotational mixing in lower metallicity stars - for more discussion see \citet{me2006}.

\begin{figure}[!ht]
\plotone{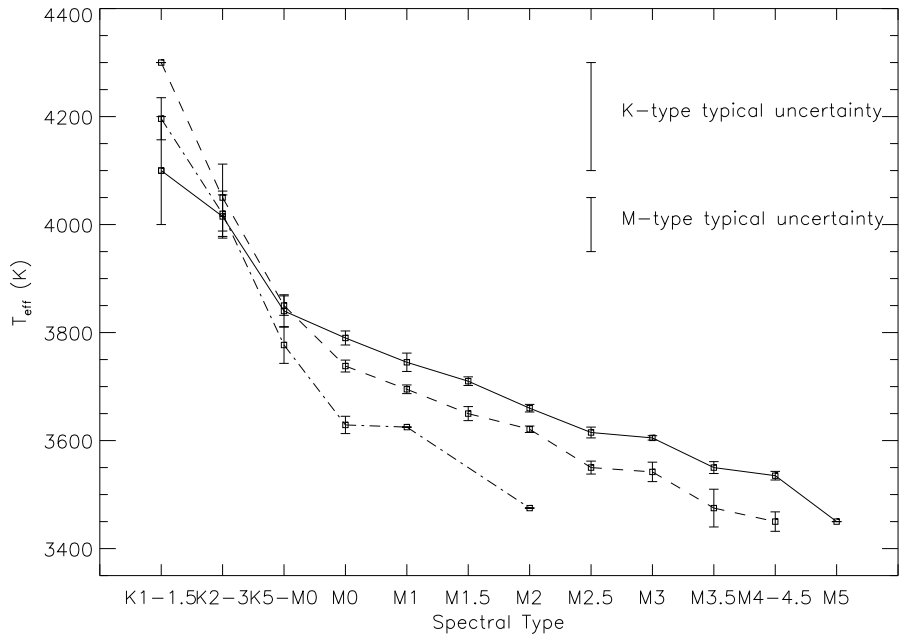}
\caption{Effective temperature scale for the Milky Way (solid line), LMC
(dashed line), and SMC (dotted-dashed line), adapted from \citet{me2006}.
The three scales together allow us to probe the effects of metallicity on
spectral type. For example, a 3650 K star would be an M2 I star in the
Milky Way, M1.5 I in the LMC, and K5-M0 I in the SMC.}
\end{figure}

\section{Late-Type Red Supergiants in the Clouds}
When comparing RSGs in the Milky Way and the Clouds, \citet{el1985} note an interesting shift in the spectral types of these stars,
with the average RSG spectral subtype shifting towards earlier types at lower metallicities. More precisely, the average RSG subtype is  found to be K5-K7 I
in the SMC, M1 I in the LMC, and M2 I in the Milky Way \citep{phil2003}. \Citet{me2006} find two distinct explanations for this
shift. The spectral subtype of late K- and M-type stars is primarily based on the strengths of the TiO bands, which are sensitive
to both effective temperature $and$ chemical abundances. We see that the average spectral subtype shifts to earlier types (and weaker
TiO) bands in the lower-metallicity environments \citep[$Z/Z_{\odot}$ = 0.2 for the SMC and $Z/Z_{\odot}$ = 0.5;][]{we1997},
suggesting that this shift in type is symptomatic of lower chemical abundances and weaker TiO lines at a given effective temperature.
This is well-illustrated in Figure 2, where we can see that a RSG with an effective temperature of 3650 K would be assigned a spectral
type of M2 I in the Milky Way, M1.5 I in the LMC, and K5-M0 I in the SMC. It is important, however, to note that this explanation
does not accomodate the full range of effects that metallicity imposes on massive stellar evolution.

Figure 10 of \citet{me2006} compares the evolutionary tracks for the Milky Way, the $Z/Z_{\odot}$ = 0.5 LMC, and the $Z/Z_{\odot}$ 0.2 SMC. There is
a clear shift of the Hayashi limit - the coolest tip of the tracks - to warmer temperatures at lower metallicity. For a 15-25
M$_{\odot}$ star we see a shift in the coolest effective temperatures for RSGs of about 100-150 K from the Milky Way to the LMC, and
about 500 K from the Milky Way to the SMC. In agreement with the speculation offered by \citet{el1985}, this does an excellent
job of explaining the observed shift in spectral subtypes. Following this explanation, one would expect that the shifting of the Hayashi
limit to warmer tempeartures at lower metallicity would therefore impose a hard limit on how cold, and hence how late-type, RSGs can be
in a particular environment.

Despite this expected restriction imposed by the Hayashi limit, there are several late-type RSGs in the Clouds that are considerably
later than the average spectral subtype, cooler than the current evolutionary tracks allow and occupying the ``forbidden'' region
to the right of the Hayashi track. We speculated that these might not all be
true members of the Clouds - instead, they could be foreground halo giants. Alternatively, they could
present a challenge to our understanding of evolutionary theory. We decided to investigate these most unusual Cloud RSGs in more
detail \citep{me2007}. We obtained moderate-resolution spectrophotometry of seven late-type RSGs in the LMC and five late-type RSGs in the SMC in December of 2005, using
the RC spectrograph on the 4-m Blanco telescope. We fit the spectra using the MARCS stellar atmosphere models
described above and probed their physical properties to better understand the nature of these 
unusually cool stars. The stars selected in our sample, and the physical properties derived from the spectral fitting, are given in
Table 1 - several stars in the sample had been previously observed in the \citet{me2006} 
November/December 2004 run but had not been published.

\subsection{Membership in the Magellanic Clouds}
One would expect that a magnitude- and color-selected sample of RSG candidates could potentially be contaminated by foreground dwarfs
or halo dwarfs and giants. Of these three, foreground dwarfs are by far the major contaminant in studying RSGs in Local Group galaxies
\citep{phil1998}, and can fortunately be easily recognized on the basis of radial velocities. \Citet{phil2003} obtained precision
radial velocities for a sample of red stars seen towards the Clouds using the CTIO 4-m telescope, and found that most of these stars
had radial velocities consistent with those of the Magellanic Clouds, excluding a small fraction (11\% for the SMC and 5.3\% for the
LMC) which had much smaller radial velocities and could be readily identified as foreground dwarfs. The remaining stars could be
tentatively identified as true RSG members of the Clouds - however, most of the apparent radial velocity of the Clouds is simply
a reflection of the sun's motion, so this RSG sample could be contaminated by red stars in the Milky Way's halo. \Citet{me2006}
estimate that this would be a few percent or less, but we reconsidered the issue when examining these late-type Cloud RSGs.

The sample of stars in \citet{me2006} and \citet{me2007} mostly have 12 $< V <$ 14, with a few fainter stars. Their $B-V$ colors are
greater than 1.6 (see table 2). According to an updated version of the \citet{ba1980} model provided by Heather Morrison,
we expect a surface density of halo giants of about 0.2 $\pm$ 0.15 deg$^{-2}$ in this magnitude/color range towards either the LMC or the
SMC. The area of the \citet{phil2002} survey was 14.2 deg$^2$ towards the LMC, and 7.2 deg$^2$ towards the SMC, so we might expect a 0.6\%
contamination by halo towards the LMC, and a 0.9\% contamination towards the SMC. We therefore expect only a fraction of a star in our
entire Cloud RSG cample of 85 stars - 73 stars from \citet{me2006} and 12 stars in \citet{me2007}. For the sample of 12
late-type RSGs, a 1\% contamination (a tenth of a star) is likely a large overestimate, given that the vast majority of halo
contaminants have $B-V$ $<$ 1.8 while all but two of our late-type stars have $B-V$ $>$ 1.8.

Finally, a more precise kinematic test can be applied to determine if the LMC stars discussed here follow the radial velocities of
other RSGs as a function of spatial position in the LMC. The kinematics of the SMC are quite complex, and so we restrict this argument
to the LMC, where the kinematics are relatively well understood \citep{ol2007}. A histogram of RSG velocities derived by \citet{ol2007}
for the LMC \citep[see Figure 1 of][]{me2007} demonstrates that the late-type LMC RSGs in our sample follow the kinematics of the
galaxy, behavior that we would not expect in the case of halo giant contaminants - for a more detailed discussion see \citet{me2007}.
We therefore conclude that our sample of late-type stars examined here is unlikely to contain foreground objects, and are in
fact true Cloud RSGs.

\subsection{Photometric Variability}
We found that all 12 of the objects in our late-typesample demonstrated large variability in their $V$ magnitudes. Using photometry
from the All Sky Automated Survey (ASAS) project \citep{po2002} along with the CCD photometry of \citet{phil2002} and values
derived from our own spectrophotometry using the $V$ band curve of \citet{be1990} and the zero-points given by \citet{be1998},,
we calculated a $\Delta V$ for each of our stars in both the 2004 and 2005 samples.
Normal RSGs are known to be variable in $V$ \citep{jo2000}, and a sample of seventy RSGs from \citet{me2006}
show an average $\Delta V$ of 0.9 mag. Each of the late-type stars described here show larger variations, averaging a $\Delta V$ of
1.3 for the SMC sample and 1.6 for the LMC sample. The most extreme case of $V$ variability is found in the case of SMC 050028,
more commonly known as HV 11423, which varies by nearly 2 mag. Many of the late-type RSG lightcurves also seem to suggest some
level of quasi-periodicity.

In contrast, we find much lower variability at $K$, which is also typical of other RSGs \citep{jo2000}; comparing the values from
2MASS and DENIS \citep{ki2004}, we find an average difference in $K$ between the two surveys of about 0.08 mag.

Finally, we find that five of our stars are unusually variable in $B-V$, with the 2005 spectrophotometry differing by several tenths
of a magnitude from the 1999/2000 values of \citet{phil2002}. $B-V$ is not very sensitive to effective temperature in RSGs
\citep[see][]{phil2007}, and we suggest that the variations we see here are instead indicative of changes in the amount of circumstellar dust,
causing differences in the reddening. As described by \citet{phil2007}, dust around RSGs results in a significant amount of
circumstellar extinction, amounting to several magnitudes in extreme cases. These variations suggest episodic dust ejection on the
timescale of a few years, consistent with the study of \citet{da1994}.

It is important to note that these dust-related changes in $B-V$ do not correlate simply with changes in the $V$ magnitudes. This
shows that the $V$-band variability is not merely symptomatic of episodic dust ejection, but is in fact a result of true physical
changes in the star. The time-resolved observations of these stars, including their $V$ magnitudes, are given in Table 2.

\subsection{Spectral Variability}
We were intrigued by the large discrepancies in spectral subtypes assigned to some of our stars by previous studies and our own past
work. While determining spectral types involves a small degree of subjectivity, there was no similar disagreement between
the \citet{phil2003} spectral types and the ones we determined for the 2004 observations from \citet{me2006}. Indeed,
spectral variability of a type or more is unheard of in RSGs.

This disparity in spectral types is most profoundly illustrated in the case of HV 11423, which also stands out as the RSG that is
the most highly variable in $V$. When we
observed the star in 2004 we assigned it an unusually early spectral type of K0-1 I, corresponding to a temperature of 4300 K. When
it was reobserved in 2005 its spectrum had changed dramatically, yielding a spectral type of M4 I and a temperature of 3500 K.
An additional spectrum of HV 11423 was obtained in September of 2006 with the CTIO 1.5-m telescope, thanks to the courtesy of F.
Walters and the SMARTS observing queue. While this spectrum did not have sufficient wavelength coverage to determine a spectral
type or effective temperature, it agreed excellently with the K0-1 I spectrum from 2004. Furthermore, a December 2001 archival
spectrum of HV 11423 from the Very Large Telescope corresponds to an even later (and hence, cooler) spectral type of M4.5-5 I, and
HV 11423 was assigned a spectral type in M0 I in both October 1978 and October 1979. The M4-5 I spectral types are by far
the latest type seen for an SMC supergiant, and the corresponding temperatures in these late-type states place HV 11423 well outside
the limits of hydrostatic equilibrium.

We also find three other stars in our sample that exhibit evidence of spectral variability between 2004 and 2005: SMC 046662 (M2 I
to K2-3 I), SMC 055188 (M2 I to M4.5 I), LMC 170452 (M4.5-5 I to M1.5 I) - from comparing the \citet{phil2003} spectra
observed in 2001 to our 2004 and 2005 observations, we can confirm that these changes are truly symptomatic of differences in
line strength, and therefore changes in the stars' physical properties (for comparison, LMC 148035 was found to vary between M4 I and
M2.5 I between 2001 and 2005, but in this case there was no compelling difference in the spectra and we concluded that the disagreement was a result of
\citet{phil2003} assigning too late a spectral type). All four of these stars - HV 11423, SMC 046662, SMC 055188, and LMC
170452 - undergo substantial changes in their effective temperature over the course of a single year, a phenomenon that has
not previously been observed in RSGs. For the three SMC stars, we can also conclude that when the stars are at their
hottest they are also more luminous (the 2004 spectrum of LMC 170452 has incomplete spectral coverage,
ranging from 4000\AA\ to 6500\AA\ as compared to the 4000\AA\ - 9000\AA\ coverage of the rest of our stars; the data in this case
is sufficient for assigning
a spectral type but insufficient for calculating physical properties such as $T_{\rm eff}$ and $A_V$). Finally, we are able to estimate the amount of extinction for the three SMC stars, and
find the $A_V$ is larger when the stars are hotter. This reddening is higher than what is typically seen for OB stars in the Clouds,
further supporting the argument that these effects are symptomatic of sporadic circumstellar dust production. Both HV 11423 and
SMC 055188 are among the only four known SMC RSGs that are IRAS sources \citep{phil2007}, suggesting that we are in fact seeing
thermal emission from these stars' circumstellar dust. These variations in the stars' physical properties are given in Table 2.
Changes in the spectra of the unusually variable SMC stars between 2004 and 2005 are illustrated in Figure 3.

\begin{figure}[!ht]
\plottwo{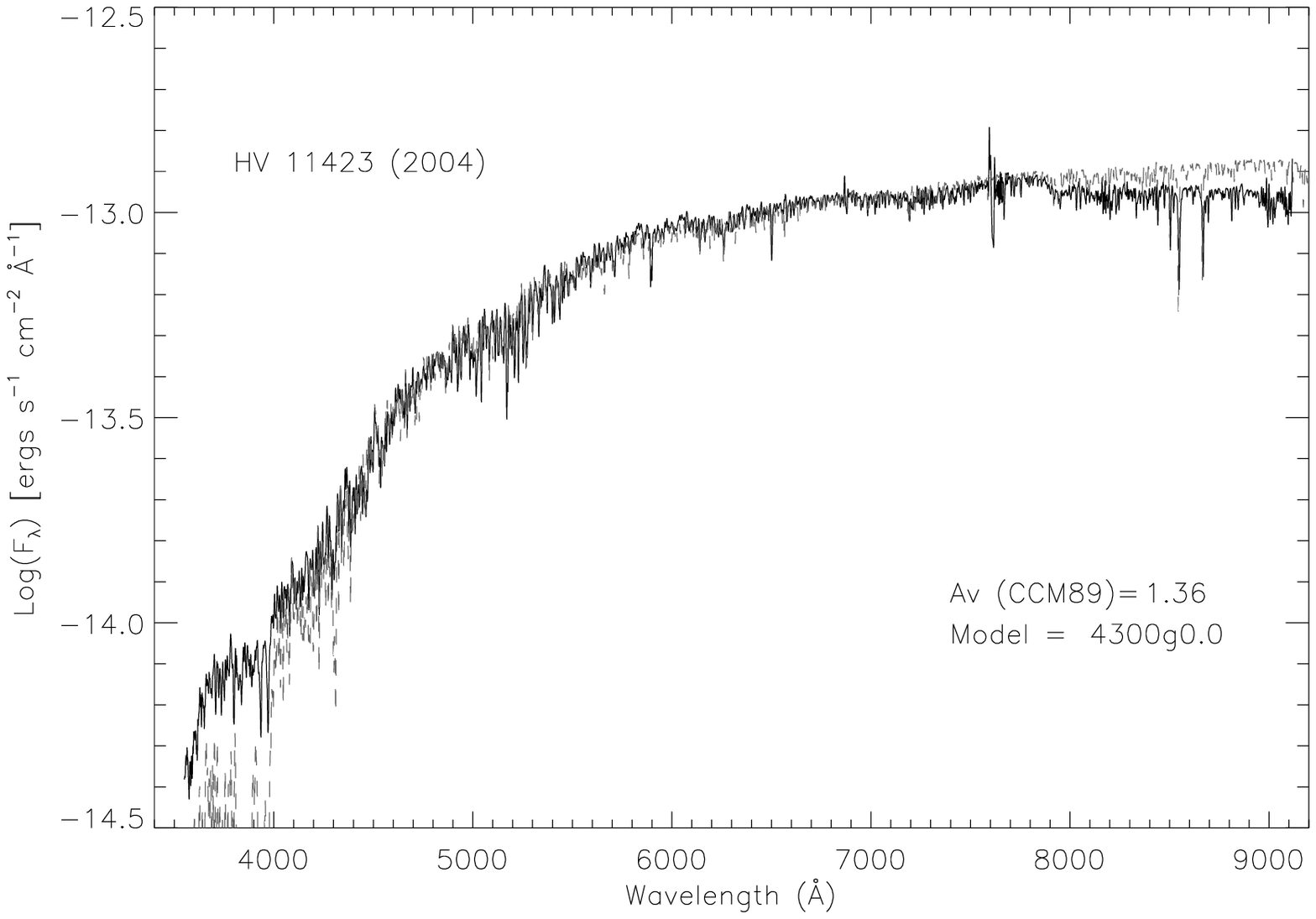}{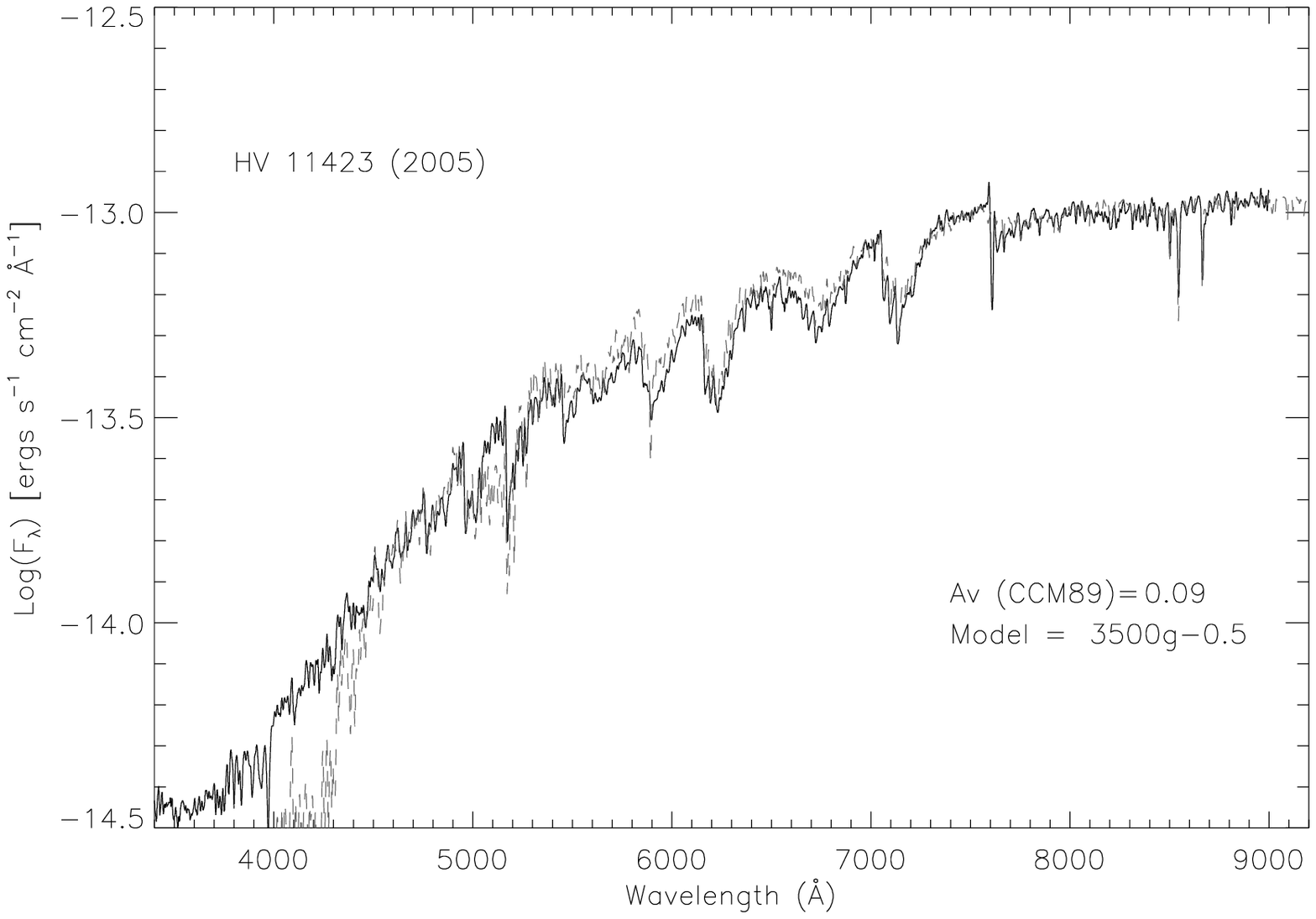}
\plottwo{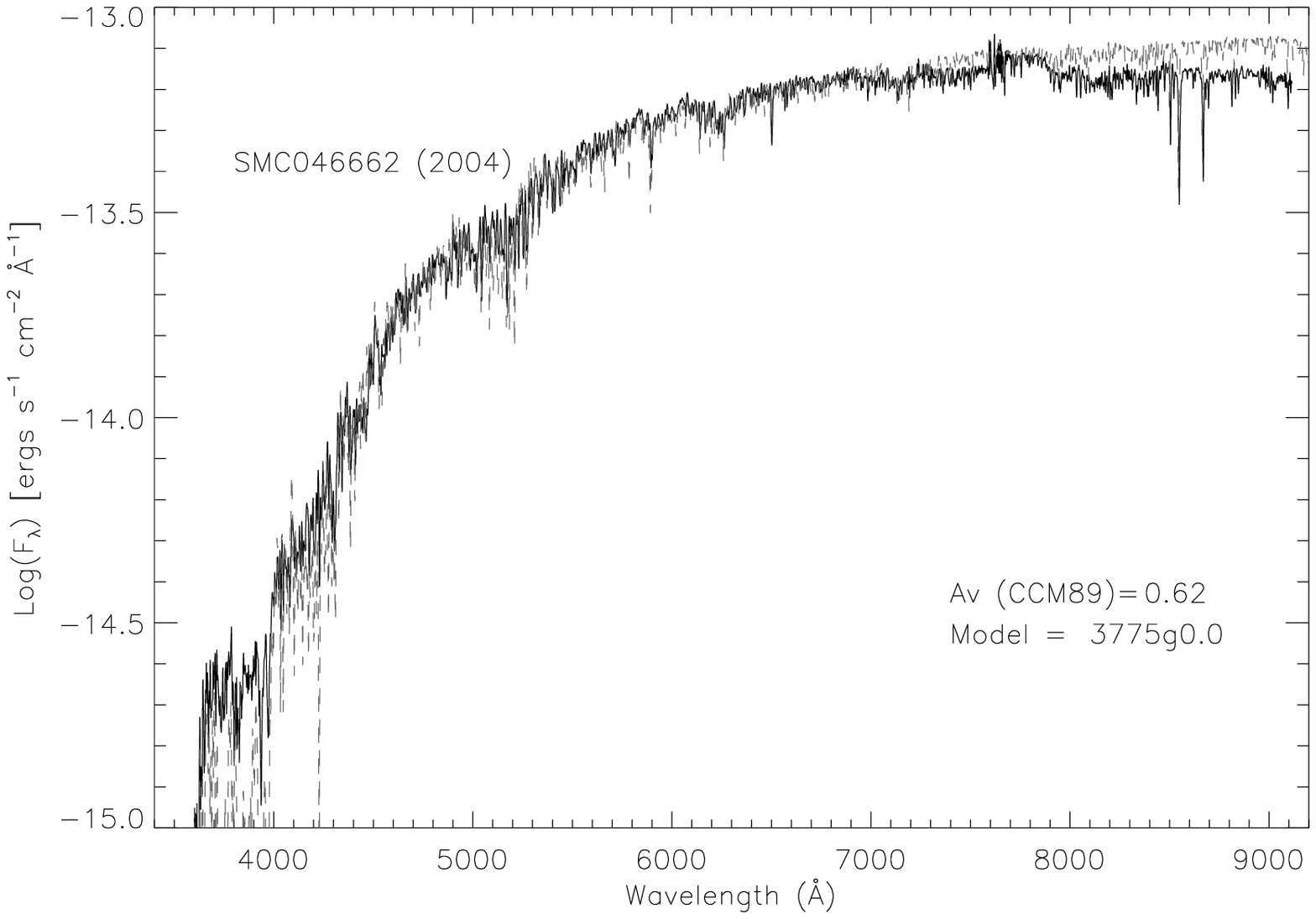}{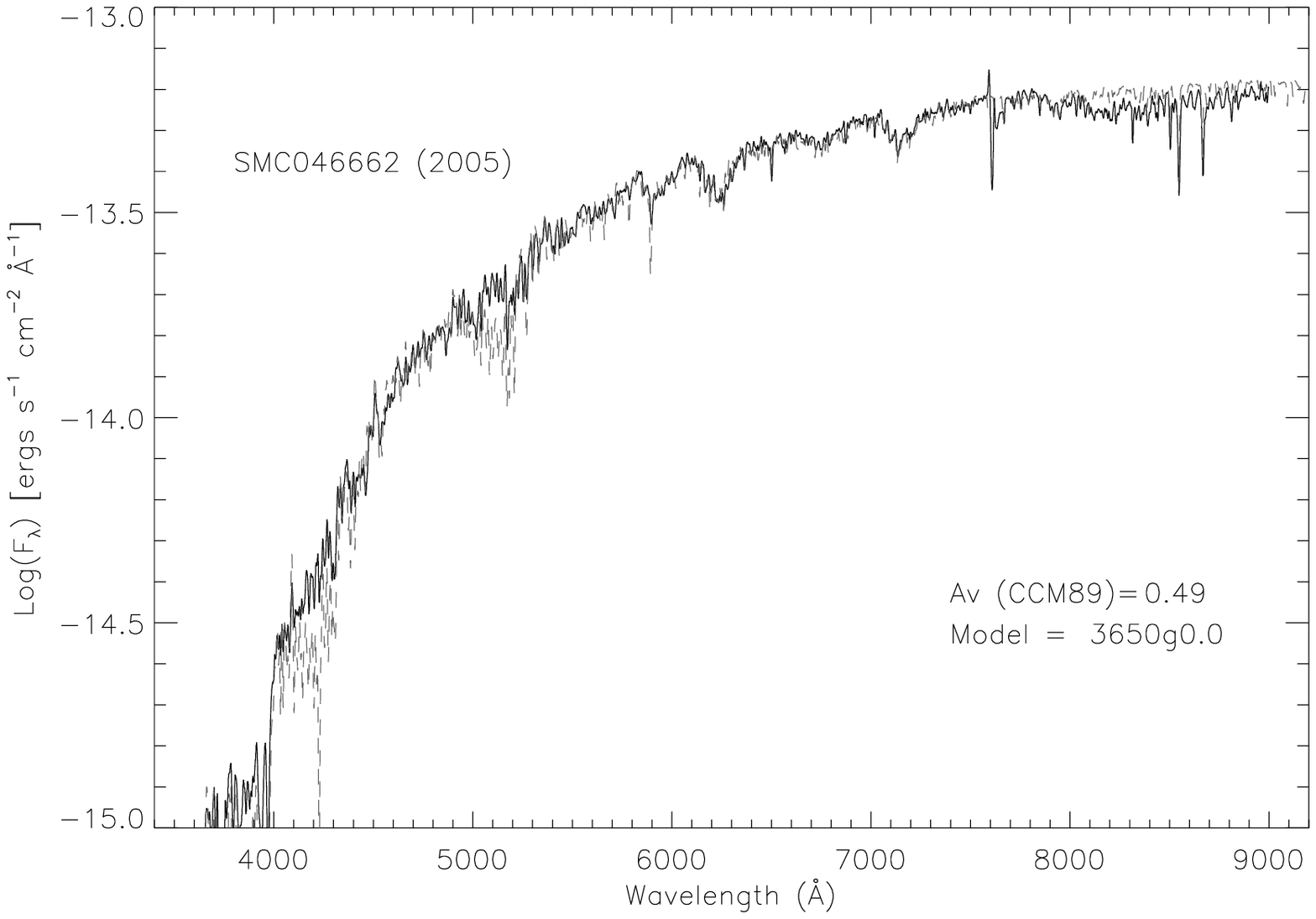}
\plottwo{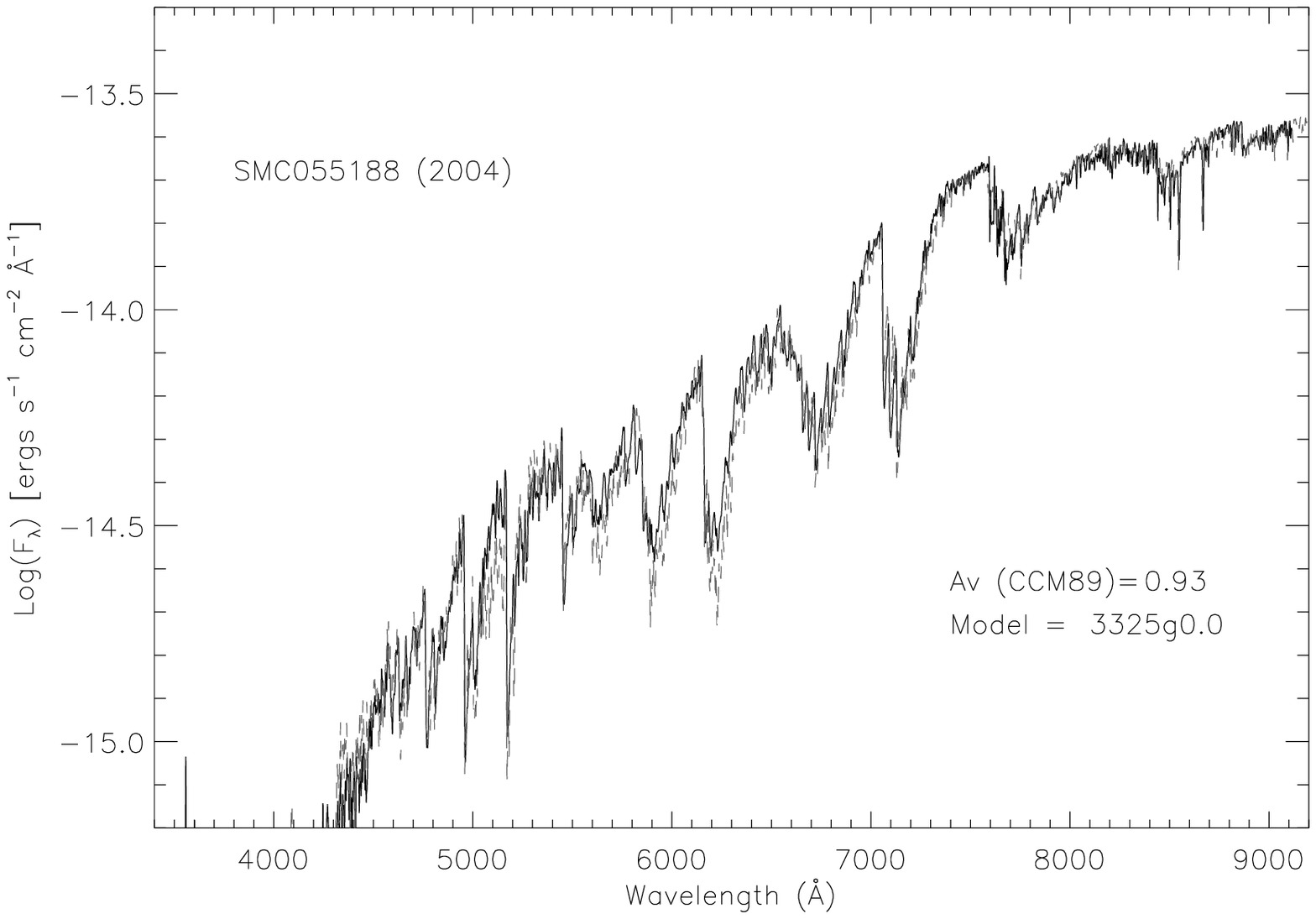}{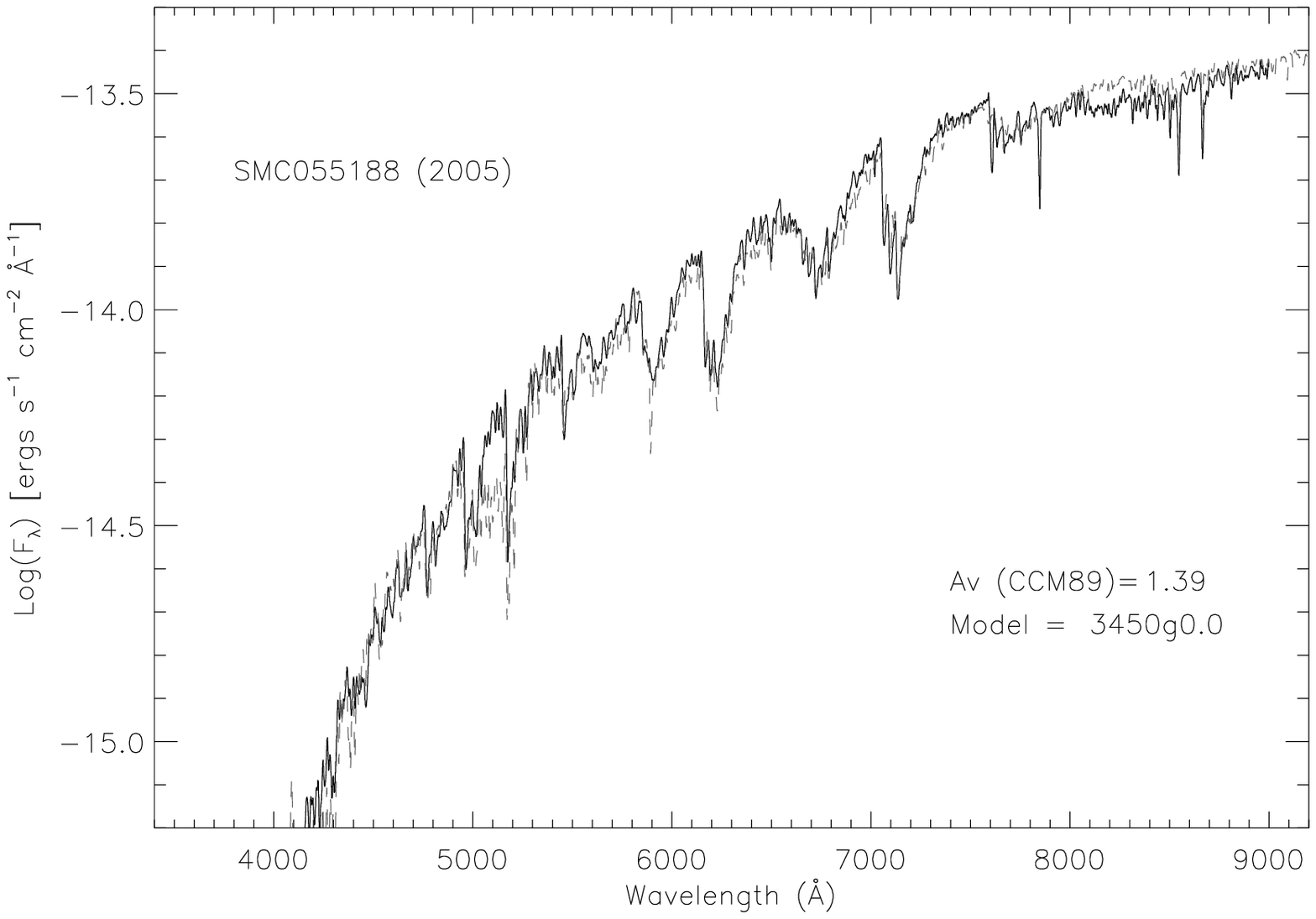}
\caption{Comparisons of our three variable SMC RSGs' spectra in 2004 (left)
and (2005) right, for HV 11423 (top), SMC 046662 (middle), and SMC 055188
(bottom). The data are shown as a solid black line, with the fits by the
reddened MARCS models overlaid as a dotted gray line and the model parameters
are given in the lower left.}
\end{figure}

\subsection{Placement on the H-R Diagram}
In Figure 4 we place our full sample of LMC and SMC late-type RSGs on the H-R diagram, accompanied by stellar evolutionary tracks
of appropriate metallicity. It is clear that stellar evolutionary theory is not in agreement with the observed parameters of these
late-type stars, with all of the stars lying on or to the right of the Hayashi limit. The location of the RSGs as derived from 
spectral fitting is, on average, 275 K coller than the tracks allow for the LMC and 541 K cooler for the SMC. When Milky Way RSGs
such as VY CMa occupied this forbidden region of the H-R diagram, the disagreement was rectified by a more accurate determination of
the stars' effective temperature or by a reevaluation of other physical properties. By contrast, it is evident in this case that the
location of these evolutionary tracks does not accommodate the full range of RSG properties and behaviors in these
lower-metallicity environments. This is further highlighted by the fact that the discrepancy is worse in the case of the SMC, where
the lower metallicity will enhance the effects of rotation on the luminosity of the evolutionary tracks \citep{ma2001} due
to the effects of mixing. Even the high-rotation (300 km s$^{-1}$) tracks are not sufficient to reproduce the location of these
stars, showing that this increased discrepancy in the SMC remains unexplained.

\begin{figure}[!ht]
\plottwo{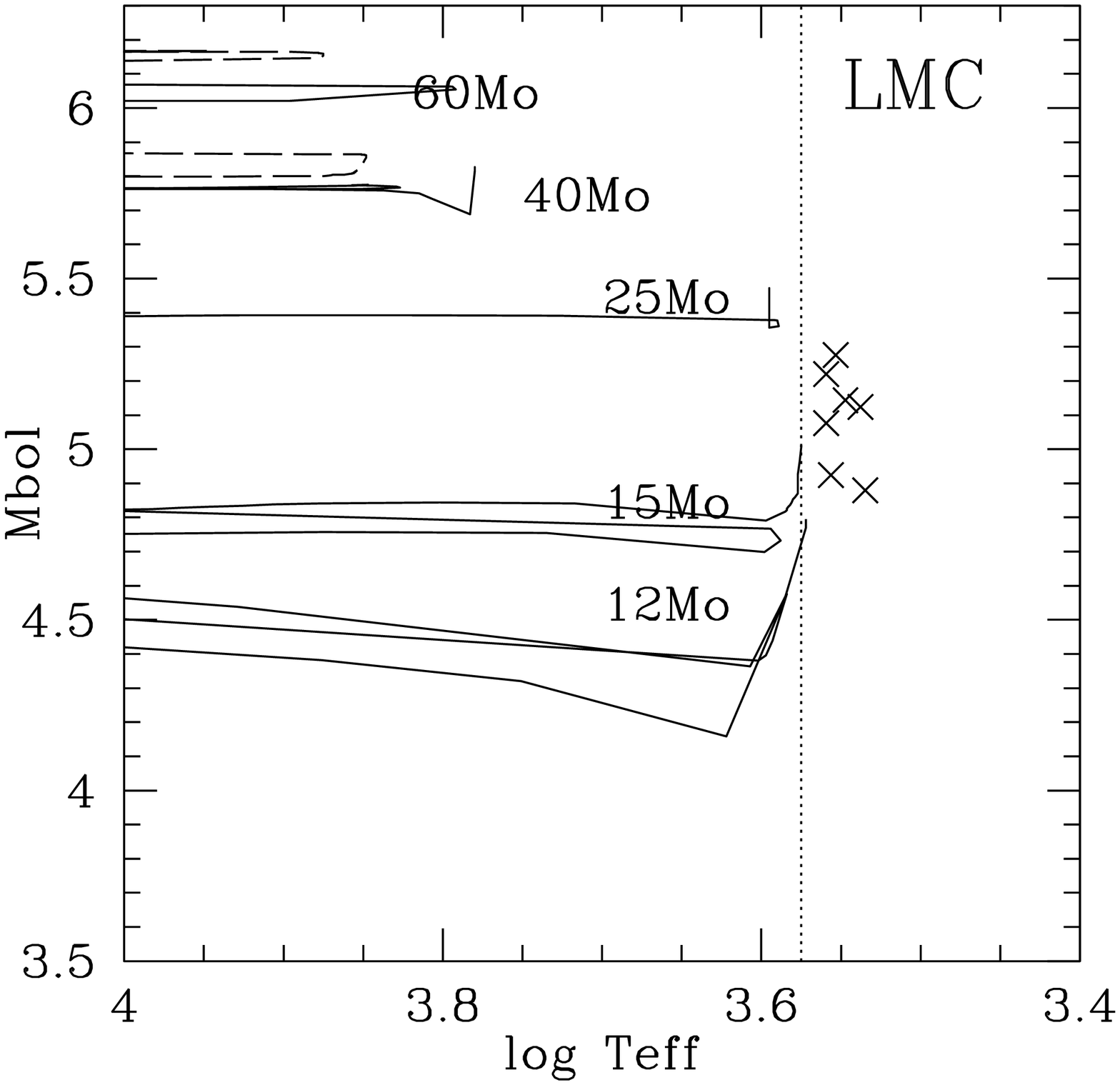}{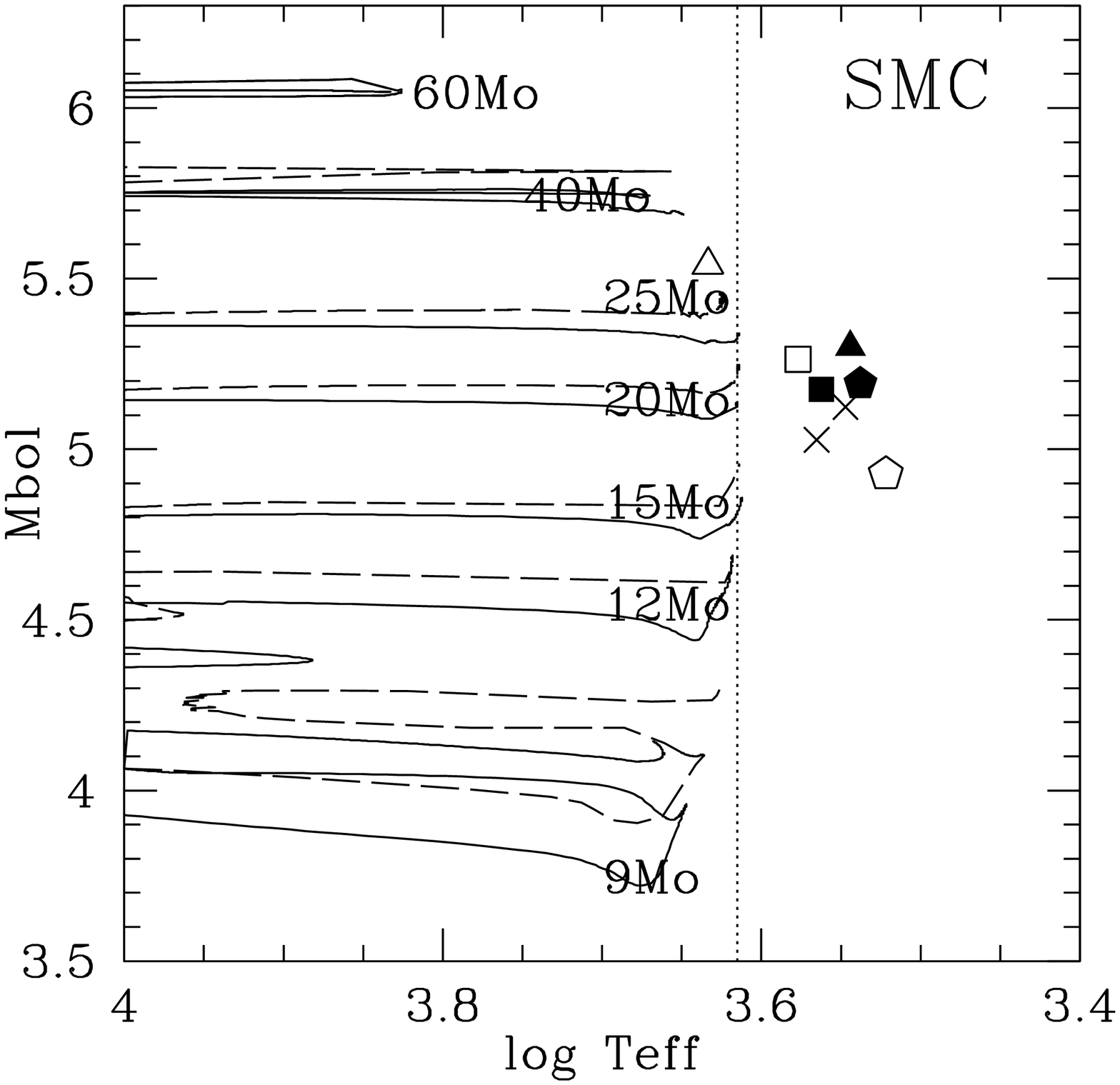}
\caption{Location of our sample of late-type MC RSGs on the H-R diagram as compared to the evolutionary. We show
the LMC (left) and SMC (right) RSGs with the physical parameters derived from fitting of the MARCS models to
the spectrophotometry. We also show evolutionary tracks of the appropriate metallicity. The non-rotation
evolutionary tracks are shown as solid black lines and come from \citet{sc1993} for the LMC and \citet{ma2001} for the SMC. The new models, which include 300 km s$^{-1}$ initial
rotation, are shown as dashed lines and come from \citet{mey2005} for the LMC and \citet{ma2001}
for the SMC. The Hayashi limit is represented by the vertical dotted line. In the case of the SMC, we use symbols
to represent the 2004 (open) and 2005 (filled) observations of our three variable SMC stars: HV 11423 (triangle),
SMC 046662 (square), and SMC 055188 (pentagon).}
\end{figure}

We also compare H-R diagrams of the late-type RSGs in the SMC in 2004 and 2005, illustrating the change in position of HV 11423,
SMC 046662, and SMC 055188. Surprisingly, we see that HV 11423 actually moves $out$ of hydrostatic equilibrium between 2004, when
it is positioned at the rightmost reaches of the evolutionary tracks, and 2005, when it has moved well into the forbidden region
to the right of the Hayashi track. All three of the stars undergo these unusual physical variations while in the Hayashi
forbidden region.

\section{What's Going On?}
HV 11423, SMC 046662, SMC 055188, and LMC 170452 all exhibit remarkably similar behavior: cool stars, inhabiting the forbidden
region to the right of the Hayashi track, that show large variability in spectral type, $V$ magnitudes, and extinction, presumed to be from episodic production
of circumstellar dust. These stars suffer changes in effective temperature and bolometric luminosity on the timescales of months; in
every case,  when they are at their hottest they are also brighter, dustier, and more luminous. These stars all occupy the Hayashi
forbidden region of the H-R diagram, where we would not expect stars to be in hydrostatic equilibrium. One would expect hydrostatic
instability to lead to this sort of variability and behavior, but the precise physical explanation for these phenomena and the
mechanisms which may lead to such an instability remains a mystery.

It is possible that such behavior is simply a case of extremes,
with the coldest stars in any environment moving to the right of the Hayashi track and undergoing these unique physical variations.
Conversely, it could be possible that such extremes are only attainable in low-metallicity environments where the Hayashi track
moves to warmer temperatures. Similarly, such behavior could simply ``ignite'' only in low-metallicity environments, a function
of the complex interrelation between massive stellar evolution and the interstellar medium environment. The question of how such
environmental factors might correlate with this behavior can be solved by observations of the latest-type Milky Way red supergiants,
as well as observations of RSGs in other low-metallicity environments and continued monitoring of these unusual stars as a means
of obtaining a more complete understanding of the physical phenomena being observed. Better knowledge of this unusual behavior could
shed new light on the nature of late-type massive star evolution, particularly at low metallicities, and highlight properties
and environmental effects that the next generation of evolutionary tracks should strive to accomodate.

\acknowledgements
We are very grateful to the staff at KPNO and CTIO for the excellent hospitality and support provided during our observations.
We acknowledge valuable assistance, correspondence and conversation about the extreme case of HV 11423 from Drs. Brian Skiff,
Beverly Smith, Geoff Clayton, and Roberto Mendez, as well as Dr. David Silva for his help in obtaining follow-up observations on
the SMARTS 1.5-m telescope. This work made use of data from the Two Micron All Sky Survey, which is a joint
project of the University of Massachusetts and the Infrared Processes and Analysis center/California Institute of Technology,
funded by the National Aeronautics and Space Administration and the National Science Foundation. This work also benefitted greatly
from the valuable resources of the All Sky Automated Survey of \citet{po2002}. This work was supported by AST-0604569. Finally,
we wish to acknowledge Drs. Don Luttermoser, Bev Smith, Rico Ignace, Gary Henson, and Mark Giroux, the organizing committee who
helped make ``The Biggest, Baddest, Coolest Stars Conference'' possible.


\begin{landscape}
\begin{table}[!ht]
\caption{Program Stars and Physical Properties}
\smallskip
\begin{center}
{\small
\begin{tabular}{lcccccccccccc}
\tableline
\noalign{\smallskip}
Star & $\alpha_{\rm 2000}$ & $\delta_{\rm 2000}$ & HJD - 2,450,000 & Type & $V$ & $M_V$ & $A_V$ & $T_{\rm eff}$ & log $g$ (Model) &$R/R_{\odot}$ &$M_{\rm bol}$ \\
\noalign{\smallskip}
\tableline
\noalign{\smallskip}

SMC 046662  &00 59 35.04 &-72 04 06.2 &3340.58 &K2-3 I  &12.38  &-7.14  &0.62 &3775 &0.0   &1000 &-8.41 \\
            &            &            &3725.63 &M0 I    &12.75  &-6.65  &0.50 &3650 &0.0   &960  &-8.19 \\
SMC 050028  &01 00 55.17 &-71 37 52.7 &3341.50 &K0 I    &11.84  &-8.40  &1.35 &4300 &0.0   &1060 &-9.10 \\
            &            &            &3724.75 &M4 I    &12.46  &-6.50  &0.10 &3500 &-0.5  &1220 &-8.5 \\
SMC 052334  &01 01 54.16 &-71 52 18.8 &3725.67 &K5-M0 I &12.87  &-6.34  &0.31 &3675 &0.0   &800  &-7.82 \\
SMC 055188  &01 03 02.38 &-72 01 52.9 &3341.59 &M4.5 I  &14.99  &-4.87  &0.96 &3325 &0.0   &870  &-7.57 \\ 
            &            &            &3725.68 &M3-4 I  &14.23  &-6.07  &1.40 &3450 &0.0   &1100 &-8.23  \\
SMC 083593  &01 30 33.92 &-73 18 41.9 &...\tablenotemark{a}     &M2 I    &12.83  &-6.16  &0.09 &3525 &0.0   &970  &-8.06 \\
LMC 143035  &05 29 03.58 &-69 06 46.3 &3725.81 &M4 I    &14.10  &-5.64  &1.24 &3450 &-0.5  &1020 &-8.06  \\
LMC 148035  &05 30 35.61 &-68 59 23.6 &3725.81 &M2.5 I  &13.42  &-6.57  &1.49 &3575 &0.0   &1130 &-8.44  \\
LMC 150040  &05 31 09.35 &-67 25 55.1 &3725.82 &M3-4 I  &13.09  &-6.03  &0.62 &3525 &-0.5  &990  &-8.11  \\
LMC 158646  &05 33 52.26 &-69 11 13.2 &3725.82 &M2 I    &13.80  &-6.25  &1.55 &3625 &-0.5  &870  &-7.94  \\
LMC 162635  &05 35 24.61 &-69 04 03.2 &3725.83 &M2 I    &14.58  &-5.78  &1.86 &3600 &0.0   &740  &-7.56  \\
LMC 168757  &05 37 36.96 &-69 29 23.5 &3727.84 &M3-4 I  &14.62  &-4.90  &1.02 &3425 &0.0   &780  &-7.45  \\
LMC 170452  &05 38 16.10 &-69 10 10.9 &3727.83 &M1.5 I  &13.99  &-6.68  &2.17 &3625 &0.0   &1060 &-8.37  \\
\noalign{\smallskip}
\tableline
\end{tabular}
}
\tablenotetext{a}{From merged 2004 and 2005 data.}
\end{center}
\end{table}
\end{landscape}

\begin{table}[!ht]
\caption{Time Resolved Data\tablenotemark{a}}
\smallskip
\begin{center}
{\small
\begin{tabular}{lcccccc}
\tableline
\noalign{\smallskip}
Star & HJD - 2,450,000 & $V$\tablenotemark{b} & $A_V$ & $B-V$ & $T_{\rm eff}$ & Spectral Type \\
\noalign{\smallskip}
\tableline
\noalign{\smallskip}
SMC 046662 &1186.58 & 12.90 &... & 1.88 &... &\... \\
&2188.56 & 13.32 &... &... &... &M2 I \\ 
&3340.58 & 12.38 &0.62 & ... &3775 & K2-3 I\\
&3725.63 & 12.75 &0.50 & 1.95 &3650 & M0 I\\
\tableline
\noalign{\smallskip}
SMC 050028 &3341.50 & 11.84 &1.35 &1.85 &4300 &K0 I \\
           &3724.75 & 12.46 &0.10 &1.93 &3500 &M4 I \\
           &4003.62 &...    &...  &...  &...  &K0 I \\
\tableline
\noalign{\smallskip}
SMC 052334 &1186.58 &   12.89 &... & 1.94 &... &... \\
&2187.56 &   12.72 &... &... &... &K7 I \\ 
&3725.67 &   12.87 &0.31 & 1.91 &3675 &K5-M0 I\\
\tableline
\noalign{\smallskip}
SMC 055188 &1186.58  & 14.96 &... & 2.25 &... &...\\
&2188.56  &... &... &... &... &M2 I \\ 
&3341.59  & 14.99 &0.96 & ... &3325 &M4.5 I\\ 
&3725.68 &  14.23 &1.40 & 2.22 &3450 &M3-4 I\\
\tableline
\noalign{\smallskip}
SMC 083593 &1186.59  & 12.64 &... & 1.87 &... &...\\
&3334.60  & 12.81 &... & ... &... &...\\ 
&3724.60\tablenotemark{c}  & 12.83 &0.09 & 1.60 &3525 &M2 I\\
\tableline
\noalign{\smallskip}
LMC 143035 &1186.67  &  13.53 &... & 1.93 &... &...\\
&1998.51  & 13.51 &... & 1.93 &... &...\\
&2188.72  & 14.24 &... &... &... &M3-4.5 I \\ 
&3334.60   & 13.99 &... & ... &... &M4.5-5 I\\
&3725.81  &  14.10 &1.24 & 1.93 &3450 &M4 I\\
\tableline
\noalign{\smallskip}
LMC 148035 &1186.67   & 12.94 &... & 1.82 &... &...\\
&1996.53   & 13.89 &... & 1.71 &... &... \\
&1998.51   & 13.87 &... & 1.62 &... &... \\
&2188.72   & 12.94 &... &... &... &M4 I \\ 
&3725.81    & 13.42 &1.49 & 1.98 &3575 &M2.5 I\\
\tableline
\noalign{\smallskip}
LMC 150040 &1186.71  & 12.86 &... & 1.97 &... &...\\
&1997.54   & 12.81 &... & 1.96  &... &...\\
&2188.82   & 12.93 &... &... &... &M4 I \\ 
&3334.60    & 12.60 &... & ... &... &M3-4 I\\ 
&3725.82   & 13.09 &0.62 & 1.90 &3525 &M3-4 I\\
\tableline
\noalign{\smallskip}
LMC 158646 &1186.71   & 12.66 &... & 2.19 &... &...\\
&1996.53    & 13.10 &... & 2.23 &... &...\\
&2188.79    & 13.54 &... &... &... &M3-4 I \\ 
&3725.82    & 13.80 &1.55 & 2.18 &3625 &M2 I\\
\tableline
\noalign{\smallskip}
LMC 162635 &1996.53   & 14.23 &... & 2.33 &... &...\\
&2188.79   & 13.52 &... &... &... &M1 I \\ 
&3725.83   & 14.58 &1.86 & 2.20 &3600 &M2 I \\
\tableline
\noalign{\smallskip}
LMC 168757 &1996.53  &  14.08 &... & 1.77 &... &...\\
&2188.76   & 13.23 &... &... &... &M3 I \\ 
&3727.84   &  14.62 &1.02 & 1.80 &3425 &M3-4 I \\
\tableline
\noalign{\smallskip}
LMC 170452 &1996.53   &  13.99 &... & 2.39 &... &...\\
&2188.76   &... &... &... &... &M4.5-5 I \\ 
&3341.70   &  15.31 &... & ... &... &M4.5-5 I \\
&3727.83    & 13.99 &2.17 & 2.24 &3625 &M1.5 I \\
\end{tabular}
}
\tablenotetext{a}{Measurements from 2,451,186-2,451,999 were obtained
using the CCD images described in \citet{phil2002}, with uncertainties of
0.01~mag or smaller. Later measurements come
from our 2004 and 2005 spectrophotometry.}
\tablenotetext{b}{The $V$ magnitudes quoted for the 2,452,188 HJD's
comes from ASAS observations on the closest approximate date
to the observations, from 2,452,184-2,452,187.}
\tablenotetext{c}{From merged 2004 and 2005 data.}
\end{center}
\end{table}

\end{document}